\documentclass[12pt]{iopart}
\usepackage{iopams}
\usepackage{graphicx}
\newlength{\figwidth}
\begin{document}

\title[Magnetic correlations around the Mott transition 
in the Kagom\'e lattice Hubbard model
]
{Magnetic correlations around the Mott transition \\
in the Kagom\'e lattice Hubbard model}

\author{T. Ohashi
\footnote[1]{
Present address: 
Condensed Matter Theory Laboratory, 
RIKEN, Wako, Saitama 351-0198, Japan \\
\indent
e-mail: t-ohashi@riken.jp
}
$^{1}$, 
S.-i. Suga$^1$, 
N. Kawakami$^1$, 
H. Tsunetsugu$^2$
}

\address{
$^1$%
Department of Applied Physics, Osaka University, 
Suita, Osaka 565-0871, Japan \\
$^2$%
Yukawa Institute for Theoretical Physics, 
Kyoto University, Kyoto 606-8502, Japan
}
%%%%% abstract %%%%%%%%%%%%%%%%%%%%%%%%%%%%%%%%%%%%%%%%%%%%%
\begin{abstract}
We study the magnetic properties around 
the Mott transition in the Kagom\'e lattice Hubbard model 
by the cellular dynamical mean field theory combined with 
quantum Monte Carlo simulations. 
By investigating the $\mathbf{q}$-dependence of the susceptibility, 
we find a dramatic change of the dominant spin fluctuations 
around the Mott transition. 
The spin fluctuations in the insulating phase 
favor down to the lowest temperature 
a spatial spin configuration in which 
antiferromagnetic correlations are strong only 
in one chain direction but almost vanishing in the others. 
\end{abstract}
%%%%% abstract %%%%%%%%%%%%%%%%%%%%%%%%%%%%%%%%%%%%%%%%%%%%%


%%%%% text %%%%%%%%%%%%%%%%%%%%%%%%%%%%%%%%%%%%%%%%%%%%%%%%%
Geometrical frustration is one of long-standing problems in spin systems. 
Recently, the frustration effects have attracted much attention 
also in itinerant electron systems. 
The observation of heavy fermion behavior in $\mathrm{LiV_2O_4}$
\cite{kondo97}, which has the pyrochlore lattice structure with a corner-sharing
network of tetrahedra, has activated 
theoretical studies of electron correlations with
geometrical frustration \cite{lacroix01,tsunetsugu02,yamashita03}. 
The discovery of superconductivity in the triangular lattice compound
$\mathrm{Na_xCoO_2 \cdot yH_2O}$ \cite{takada03}
and the $\beta$-pyrochlore osmate $\mathrm{KOs_2O_6}$ \cite{yonezawa04}
has further stimulated intensive studies of frustrated electron systems. 
Geometrical frustration has uncovered new aspects
of the Mott metal-insulator transition.
Among others, a novel quantum liquid ground state was suggested 
for the insulating phase of the triangular lattice \cite{kashima01}, 
and this may be relevant for frustrated organic materials such as 
$\kappa$-$\mathrm{(ET)_2Cu_2(CN)_3}$ \cite{shimizu03}.

\begin{figure}[bt]
\begin{center}
\includegraphics[clip,width=0.6\textwidth]{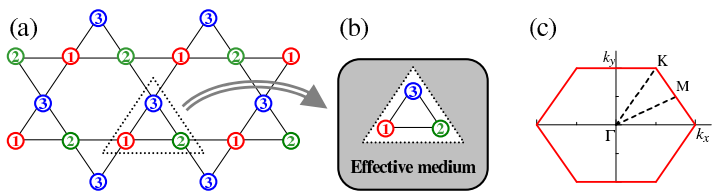}
\end{center}
\caption{
(a) Sketch of the Kagom\'e lattice and 
(b) the effective cluster model using three-site cluster CDMFT. 
(c) First Brillouin zone of the Kagom\'e lattice. 
}
\label{fig:kagome}
\end{figure}

The Kagom\'e lattice (Fig. \ref{fig:kagome}) is another prototype of 
frustrated systems showing many essential properties with the pyrochlore lattice. 
It is suggested that a correlated electron system on the Kagom\'e lattice 
can be an effective model of $\mathrm{Na_xCoO_2 \cdot yH_2O}$ 
by properly considering anisotropic hopping matrix elements 
in the cobalt 3$d$ orbitals \cite{koshibae03}. 
The electron system on the Kagom\'e lattice in the metallic regime 
was studied recently by using the fluctuation exchange (FLEX) approximation \cite{imai03} 
and quantum Monte Carlo (QMC) method \cite{bulut05}, etc \cite{lauchli04,indergand06}. 
In our recent paper \cite{ohashi06}, 
we have studied electron correlations in 
the Kagom{\'e} lattice Hubbard model, 
and found the first-order Mott transition 
at the Hubbard interaction $U/W \sim 1.37$ ($W$: band width). 

In this paper, we focus on the magnetic properties of 
the Kagom{\'e} lattice Hubbard model. 
By applying the cellular dynamical mean field theory (CDFMT) 
\cite{kotliar01}, 
we discuss the effects of geometrical frustration around 
the metal-insulator transition where 
frustration is stronger than in the weak coupling regime. 
%
%%%%%%%%%%%%%%%%%%%%%%%%%%%%%%%%%%%%%%%%%%%%%%%%%%%%%%%%%%%%
%%%%%%%%%%%%%%%%%%%%%%%%%%%%%%%%%%%%%%%%%%%%%%%%%%%%%%%%%%%%
%
We consider the standard Hubbard model with nearest-neighbor  
hopping $t>0$ on the Kagom\'e lattice,
%%%%%%%%%%%%%%%%%%%%%%%%%%%%%%%%%%%%%%%%%%%%%%%%%%%%%%%%%%%%
\begin{eqnarray}
H= - t \sum_{\left \langle i,j \right \rangle ,\sigma}
c_{i\sigma }^\dag c_{j\sigma}
+ U \sum_{i} n_{i\uparrow} n_{i\downarrow} ,
\label{eqn:hm}
\end{eqnarray}
%%%%%%%%%%%%%%%%%%%%%%%%%%%%%%%%%%%%%%%%%%%%%%%%%%%%%%%%%%%%
with $n_{i\sigma}=c_{i\sigma}^\dag c_{i\sigma}$, 
where $c_{i\sigma }^\dag$ ($c_{j\sigma}$) creates (annihilates) 
an electron with spin $\sigma$ at the site $i$.
In the following, we use the band width $W=6t$ as the energy unit. 
The dynamical mean field theory (DMFT) 
\cite{georges96} has given substantial theoretical 
progress in the field of the Mott transition 
but it does not incorporate spatially extended correlations.
Therefore in order to take account of geometrical frustration, 
we use CDMFT, a cluster extension of DMFT \cite{kotliar01,biroli04,maier05}, 
which has been successfully applied to frustrated systems such as 
the Hubbard model on the triangular lattice \cite{parcollet04,civelli05,kyung06}. 

In CDMFT, the original lattice is regarded 
as a superlattice consisting of clusters, which 
is then mapped onto an effective cluster model via a standard DMFT procedure. 
As shown in Fig. \ref{fig:kagome}, the Kagom{\'e} lattice Hubbard model 
is mapped onto a three-site cluster coupled to 
the self-consistently determined medium, 
%%%%%%%%%%%%%%%%%%%%%%%%%%%%%%%%%%%%%%%%%%%%%%%%%%%%%%%%%%%
\begin{eqnarray}
\hspace{-2cm}
S_\mathrm{eff} = 
\int_0^\beta d\tau d\tau' \sum_{\gamma,\delta,\sigma}
c_{\gamma\sigma}^\dag \left( \tau \right)
\mathcal{G}_{\gamma \delta \sigma}^{-1} 
\left( \tau - \tau' \right)
c_{\delta\sigma} \left( \tau' \right)
+ U \int_0^\beta d\tau 
\sum_{\gamma}
n_{\gamma \uparrow} \left( \tau \right) 
n_{\gamma \downarrow} \left( \tau \right). 
\label{eqn:action}
\end{eqnarray}
%%%%%%%%%%%%%%%%%%%%%%%%%%%%%%%%%%%%%%%%%%%%%%%%%%%%%%%%%%%%
Given the Green's function for the effective medium, 
$\hat{\mathcal{G}}_{\sigma}$, 
we can compute the cluster Green's function $\hat{G}_{\sigma}$
and the cluster self-energy $\hat{\Sigma}_{\sigma}$ 
by solving the effective cluster model with QMC method \cite{hirsch86}. 
Here, $\hat{\mathcal{G}}_{\sigma}$, $\hat{G}_{\sigma}$, 
and $\hat{\Sigma}_{\sigma}$ are described by 
$3 \times 3$ matrices. 
The effective medium $\hat{\mathcal{G}}_{\sigma}$ 
is then computed via the Dyson equation, 
%%%%%%%%%%%%%%%%%%%%%%%%%%%%%%%%%%%%%%%%%%%%%%%%%%%%%%%%%%%%
\begin{eqnarray}
\hat{\mathcal{G}}_{\sigma}^{-1} \left( \omega \right) &=& 
\left[ \sum_\mathbf{k} \hat{g}_\sigma \left( \mathbf{k}: \omega \right) \right]^{-1}
+ \hat{\Sigma}_{\sigma} \left( \omega \right) , 
\label{eqn:bath} \\
\hat{g}_\sigma \left( \mathbf{k} : \omega \right) &=&
\left[ \omega + \mu - \hat{t} \left( \mathbf{k} \right)
- \hat{\Sigma}_\sigma \left( \omega \right) \right ]^{-1}, 
\label{eqn:green}
\end{eqnarray}
%%%%%%%%%%%%%%%%%%%%%%%%%%%%%%%%%%%%%%%%%%%%%%%%%%%%%%%%%%%%
where $\mu$ is the chemical potential and 
$\hat{t} \left( \mathbf{k} \right)$ is the 
Fourier-transformed hopping matrix for the superlattice, 
%%%%%%%%%%%%%%%%%%%%%%%%%%%%%%%%%%%%%%%%%%%%%%%%%%%%%%%%%%%%
\begin{eqnarray}
t_{\gamma \delta} \left( \mathbf{k} \right) =
\sum_{i,j} e^{-\mathbf{k} \cdot (\mathbf{r}_i - \mathbf{r}_j)}
t_{\gamma \delta} \left( i,j \right). 
\label{eqn:hopping}
\end{eqnarray}
%%%%%%%%%%%%%%%%%%%%%%%%%%%%%%%%%%%%%%%%%%%%%%%%%%%%%%%%%%%%
Here the summation of $\mathbf{k}$ is taken over the reduced Brillouin zone 
of the superlattice (see Fig. \ref{fig:kagome}(c)). 
After twenty iterations of this procedure, 
numerical convergence is reached. In each iteration, 
we typically use $10^6$ QMC sweeps and 
Trotter time slices $L = 2W/T$ 
to reach sufficient computational accuracy. 
Furthermore, 
we exploit an interpolation scheme based on a high-frequency expansion
of the discrete imaginary-time Green's function obtained by QMC 
\cite{oudovenko02} in order to reduce time slice errors. 

We now investigate the magnetic correlation around 
the Mott transition in the Kagom{\'e} lattice Hubbard model. 
We calculate the wavevector-dependence of 
the static susceptibility, 
%%%%%%%%%%%%%%%%%%%%%%%%%%%%%%%%%%%%%%%%%%%%%%%%%%%%%%%%%%%%
\begin{eqnarray}
\chi_{\gamma \delta}(\mathbf{q}) = \int_0^{1/T} d \tau 
\sum_{\mathbf{k},\mathbf{k}'} 
\left \langle 
c_{\mathbf{k}\gamma\uparrow}^\dag               \left ( \tau \right ) 
c_{\mathbf{k}+\mathbf{q}\gamma\downarrow}       \left ( \tau \right )
c_{\mathbf{k}'+\mathbf{q}\delta\downarrow}^\dag \left ( 0 \right )
c_{\mathbf{k}'\delta\uparrow}                   \left ( 0 \right )
\right \rangle ,
\end{eqnarray}
%%%%%%%%%%%%%%%%%%%%%%%%%%%%%%%%%%%%%%%%%%%%%%%%%%%%%%%%%%%%
where $\gamma,\delta =1,2,3$ denote the site indices in the unit cell. 
We employ the standard procedure in DMFT to calculate 
$\chi_{\gamma \delta}(\mathbf{q})$ \cite{georges96}, 
which includes nearest-neighbor correlations as well as on-site correlations. 
In order to obtain $\chi_{\gamma \delta}(\mathbf{q})$,
we first calculate the two-particle Green's function  
in the effective cluster model (\ref{eqn:action}), 
%%%%%%%%%%%%%%%%%%%%%%%%%%%%%%%%%%%%%%%%%%%%%%%%%%%%%%%%%%%%
\begin{eqnarray}
C_{\gamma \delta} \left( i\omega_l,i\omega_m \right) &=& T
\int_0^\beta \int_0^\beta \int_0^\beta \int_0^\beta
d\tau_1 d\tau_2 d\tau_3 d\tau_4 \nonumber \\
&\times&
e^{-i\omega_l \left( \tau_1 - \tau_2 \right)}
e^{-i\omega_m \left( \tau_3 - \tau_4 \right)}
C_{\gamma \delta} \left( \tau_1,\tau_2,\tau_3,\tau_4 \right), \\
C_{\gamma \delta} \left( \tau_1,\tau_2,\tau_3,\tau_4 \right) &=& 
\left \langle \mathrm{T}_\tau
c_{\gamma\uparrow}^\dag   \left ( \tau_1 \right ) 
c_{\gamma\downarrow}      \left ( \tau_2 \right )
c_{\delta\downarrow}^\dag \left ( \tau_3 \right )
c_{\delta\uparrow}        \left ( \tau_4 \right )
\right \rangle ,
\label{eqn:cf}
\end{eqnarray}
%%%%%%%%%%%%%%%%%%%%%%%%%%%%%%%%%%%%%%%%%%%%%%%%%%%%%%%%%%%%
and extract the vertex function 
$\Gamma_{\gamma\delta}\left( i\omega_l,i\omega_m \right)$ 
via the Bethe-Salpeter equation, 
%%%%%%%%%%%%%%%%%%%%%%%%%%%%%%%%%%%%%%%%%%%%%%%%%%%%%%%%%%%%
\begin{eqnarray}
\hat{\Gamma} = {\hat{C^0}}^{-1} - \hat{C}^{-1}, 
\label{eqn:vf}
\end{eqnarray}
%%%%%%%%%%%%%%%%%%%%%%%%%%%%%%%%%%%%%%%%%%%%%%%%%%%%%%%%%%%%
where $C^0$ is the bare two-particle Green's function, 
%%%%%%%%%%%%%%%%%%%%%%%%%%%%%%%%%%%%%%%%%%%%%%%%%%%%%%%%%%%%
\begin{eqnarray}
C^0_{\gamma\delta}(i\omega_l) = - \frac{1}{T}
\left[ \sum_\mathbf{k} g_{\gamma\delta\downarrow} 
\left( \mathbf{k}: i\omega_l \right) \right]
\left[ \sum_\mathbf{k} g_{\delta\gamma\uparrow} 
\left( \mathbf{k}: i\omega_l \right) \right]. 
\label{eqn:cf0}
\end{eqnarray}
%%%%%%%%%%%%%%%%%%%%%%%%%%%%%%%%%%%%%%%%%%%%%%%%%%%%%%%%%%%%
On the other hand, 
the bare $\mathbf{q}$-dependent Green's function in the lattice system 
is calculated by 
%%%%%%%%%%%%%%%%%%%%%%%%%%%%%%%%%%%%%%%%%%%%%%%%%%%%%%%%%%%%
\begin{eqnarray}
C^0_{\gamma\delta}(\mathbf{q}:i\omega_l) = - \frac{1}{T} \sum_\mathbf{k} 
g_{\gamma\delta\downarrow} \left( \mathbf{k}+\mathbf{q}: i\omega_l \right)
g_{\gamma\delta\uparrow}   \left( \mathbf{k}           : i\omega_l \right). 
\label{eqn:cf0q}
\end{eqnarray}
%%%%%%%%%%%%%%%%%%%%%%%%%%%%%%%%%%%%%%%%%%%%%%%%%%%%%%%%%%%%
By using Eqs. (\ref{eqn:vf}) and (\ref{eqn:cf0q}), 
we can compute the lattice $\mathbf{q}$-dependent Green's function, 
%%%%%%%%%%%%%%%%%%%%%%%%%%%%%%%%%%%%%%%%%%%%%%%%%%%%%%%%%%%%
\begin{eqnarray}
\hat{C}\left( \mathbf{q} \right) = 
\left[ {\hat{C^0}\left( \mathbf{q} \right)}^{-1} - \hat{\Gamma} \right]^{-1}. 
\label{eqn:cfq}
\end{eqnarray}
%%%%%%%%%%%%%%%%%%%%%%%%%%%%%%%%%%%%%%%%%%%%%%%%%%%%%%%%%%%%
Taking account of the phase factor, 
we finally obtain the $\mathbf{q}$-dependent susceptibility, 
%%%%%%%%%%%%%%%%%%%%%%%%%%%%%%%%%%%%%%%%%%%%%%%%%%%%%%%%%%%%
\begin{eqnarray}
\chi_{\gamma \delta}(\mathbf{q}) =
T^2 \sum_{l,m} C_{\gamma\delta} \left( \mathbf{q}:i\omega_l,i\omega_m \right)
e^{-i\mathbf{q} \cdot \left( \mathbf{r}_\gamma - \mathbf{r}_\delta \right)}. 
\label{eqn:chi}
\end{eqnarray}
%%%%%%%%%%%%%%%%%%%%%%%%%%%%%%%%%%%%%%%%%%%%%%%%%%%%%%%%%%%%

%%%%%%%%%%%%%%%%%%%%%%%%%%%%%%%%%%%%%%%%%%%%%%%%%%%%%%%%%%%%
\begin{figure}[bt]
\begin{center}
\includegraphics[clip,width=\textwidth]{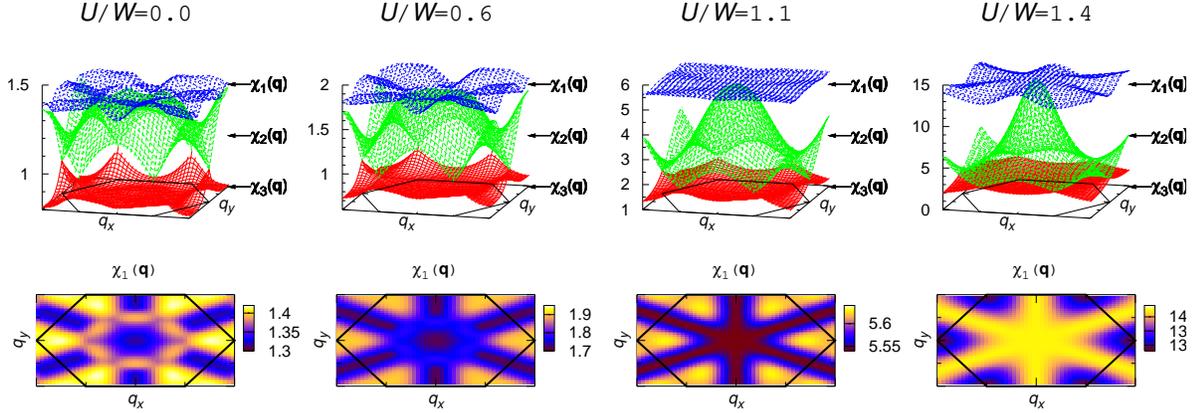}
\end{center}
\caption{
The wavevector dependence of the static susceptibility 
$\chi_m(\mathbf{q})$ for several values of $U/W$ at $T/W=1/30$. 
The three dimensional plots of $\chi_m(\mathbf{q})$ are shown 
in upper panels, from top to bottom $m=1$, $2$, $3$. 
The two dimensional plots in the lower panels 
show the dominant mode of the susceptibility $\chi_1 (\mathbf{q})$
in upper panels. 
Hexagons in figures denotes the first Brillouin zone as shown 
Fig. \ref{fig:kagome} (c). 
}
\label{fig:sus}
\end{figure}
%%%%%%%%%%%%%%%%%%%%%%%%%%%%%%%%%%%%%%%%%%%%%%%%%%%%%%%%%%%%

%%%%%%%%%%%%%%%%%%%%%%%%%%%%%%%%%%%%%%%%%%%%%%%%%%%%%%%%%%%%
\begin{figure}[bt]
\begin{center}
\includegraphics[clip,width=\textwidth]{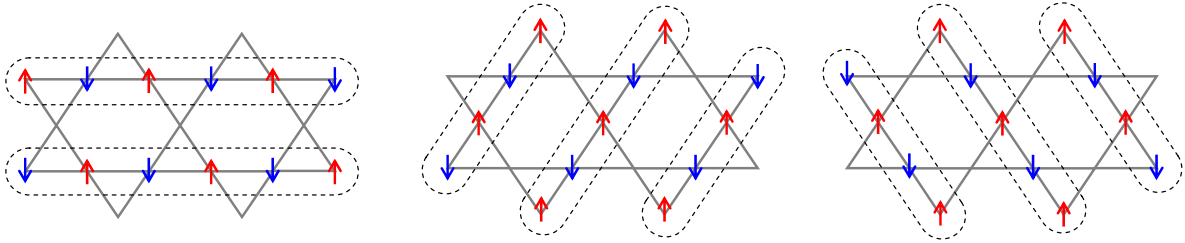}
\end{center}
\caption{
The enhanced spin correlations 
in the insulating phase $U/W=1.4$ at $T/W=1/30$. 
}
\label{fig:con}
\end{figure}
%%%%%%%%%%%%%%%%%%%%%%%%%%%%%%%%%%%%%%%%%%%%%%%%%%%%%%%%%%%%

It is convenient to introduce $\chi_m(\mathbf{q})$ for three normal 
modes ($m=1,2,3$) by diagonalizing the $3 \times 3$ matrix 
$\chi_{\gamma \delta}(\mathbf{q})$. 
In the upper panels of Fig. \ref{fig:sus}, 
we show the three eigenmodes of the susceptibility 
at $T/W=1/30$ for several values of interaction strength $U/W$. 
In the noninteracting case, 
the largest eigenvalue of the susceptibility $\chi_1(\mathbf{q})$
takes a maximum at six points in the Brillouin zone, 
the second largest one $\chi_2(\mathbf{q})$ has a maximum 
at $\mathbf{q}=(0,0)$, 
and the smallest one $\chi_1(\mathbf{q})$ takes maxima 
at the corners of the Brillouin zone. 
As $U/W$ increases, localized moments are formed and 
the susceptibility is enhanced as expected. 
In particular, 
the $\mathbf{q}=(0,0)$ peak of $\chi_2(\mathbf{q})$ becomes strongly enhanced, 
which is consistent with the previous QMC study \cite{bulut05}. 
On the other hand, the dominant mode of the susceptibility $\chi_1(\mathbf{q})$
shows only weak $\mathbf{q}$-dependence. 
As $U/W$ increases, $\chi_1(\mathbf{q})$ is enhanced not only at 
the six points above mentioned but also enhanced 
on the lines passing through $\mathrm{\Gamma}$ and $\mathrm{M}$ points, 
so that the $\mathbf{q}$-dependence of the susceptibility gets suppressed 
and becomes much weaker at $U/W=1.1$ than in the noninteracting case. 
This behavior is consistent with the previous FLEX calculation 
in the weak coupling regime \cite{imai03}. 
We confirm that the feature of the suppressed $\mathbf{q}$-dependence of 
the dominant magnetic mode 
due to geometrical frustration persists up to fairly large-$U$ region. 

%%%%%%%%%%%%%%%%%%%%%%%%%%%%%%%%%%%%%%%%%%%%%%%%%%%%%%%%%%%%
\begin{figure}[bt]
\begin{center}
\includegraphics[clip,width=\figwidth]{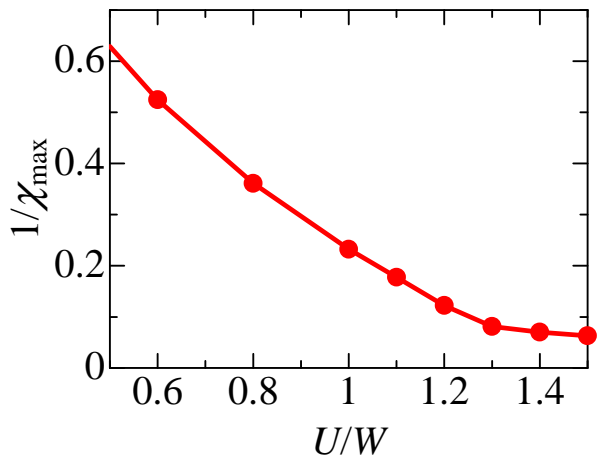}
\end{center}
\caption{
The inverse of the maximum susceptibility $1/\chi_\mathrm{max}$ 
as a function of $U/W$ at $T/W=1/30$. 
}
\label{fig:max}
\end{figure}
%%%%%%%%%%%%%%%%%%%%%%%%%%%%%%%%%%%%%%%%%%%%%%%%%%%%%%%%%%%%

We further find notable results in the insulating phase. 
The Mott metal-insulator transition occurs at $U/W \sim 1.37$ \cite{ohashi06}. 
As shown in the lower panels of Fig. \ref{fig:sus}, 
once the system enters the insulating phase, 
the $\mathbf{q}$-dependence of $\chi_1(\mathbf{q})$ dramatically changes 
its character due to the enhancement of short range 
antiferromagnetic (AF) correlations \cite{ohashi06}. 
At $U/W=1.4$, the susceptibility takes the maximum value along 
the three lines in $\mathbf{q}$ space instead of the six points 
in the weak coupling regime. 
Furthermore, by investigating the eigenvectors of $\chi_1(\mathbf{q})$, 
we find that two spins in the unit cell are antiferromagnetically coupled 
but the other spin is free. 
Therefore, at these temperatures, 
the enhanced spin fluctuations 
favor a spatial spin configuration in which 
one-dimensional (1D) AF-correlated spin chains 
are independently formed in three distinct directions. 
The three types of enhanced spin correlations are illustrated in Fig. \ref{fig:con}. 
This spin correlation is one of the naturally expected 
spin correlation on the Kagom{\'e} lattice, 
because it stabilizes antiferromagnetic configurations in one direction,
which is more stable than the naively expected spin configuration 
having a singlet pair and a free spin in each cluster. 
These 1D correlations in the finite-$T$ Mott insulating phase 
are different from the results for the Heisenberg model 
on the Kagom{\'e} lattice with the nearest-neighbor exchange 
obtained by both classical and semi-classical 
approximations \cite{chubkov92,harris92}, 
but are similar to the $\mathbf{q}=0$ structure predicted 
for the classical Heisenberg model 
with a further neighbor exchange \cite{harris92}. 
The essential difference from Ref. \cite{harris92} 
is that there is almost no correlation 
between the different chains in our results for the Hubbard model. 
It remains an interesting problem to compare the q-dependence of 
the susceptibility in the large-$U$ and low-$T$ regime with 
the results for the Heisenberg model. 

Finally, we show the maximum value of the susceptibility $\chi_\mathrm{max}$ 
as a function of $U/W$ in Fig. \ref{fig:max}. 
As $U/W$ increases towards the Mott transition point, 
$1/\chi_\mathrm{max}$ decreases almost linearly as expected. 
When $U/W$ is further increased and the system enters the insulating phase, 
the spin correlation changes into the 1D one as discussed above, 
and then $\chi_\mathrm{max}$ saturates. 
Therefore, we find no evidence of the real instability to 
an 1D ordering in the present calculation.
However, such enhanced spin fluctuations affect low-energy dynamics 
in the insulating phase \cite{ohashi06}. 

In summary, we have studied the magnetic properties around 
the Mott transition in the Kagom\'e lattice Hubbard model 
by means of CDMFT combined with QMC. 
We have investigated the $\mathbf{q}$-dependence of the susceptibility, 
and obtained consistent results with the previous studies: 
the second-largest eigenmode of the susceptibility 
shows the strong $\mathbf{q}$-dependence, 
taking a maximum at $\mathbf{q}=(0,0)$, and
the $\mathbf{q}$-dependence of the maximum eigenmode 
of the susceptibility becomes suppressed 
as $U$ increases in the metallic phase. 
We also find a dramatic change of the dominant spin fluctuations 
around the Mott transition. 
The spin fluctuations in the insulating phase 
favor a novel spatial spin configuration in which 
1D AF-correlated spin chains are independently formed 
in three distinct directions. 
Although the spin liquid state or other nonmagnetic ordered states 
may be stabilized at zero temperature \cite{misguich04}, 
the enhanced 1D spin correlations could emerge 
in the finite-$T$ Mott insulating phase. 

%%%%% acknowledgment %%%%%%%%%%%%%%%%%%%%%%%%%%%%%%%%%%%%%%%
The authors thank Y. Motome, A. Koga, and Y. Imai for valuable discussions. 
A part of numerical computations was done at the Supercomputer Center 
at ISSP, University of Tokyo and also at YITP, Kyoto university. 
This work was partly supported by Grant-in-Aid for Scientific Research
(No. 16540313) and for Scientific Research on Priority Areas
(No. 17071011 and 18043017) from MEXT. 
%%%%%%%%%%%%%%%%%%%%%%%%%%%%%%%%%%%%%%%%%%%%%%%%%%%%%%%%%%%%

%%%%%%%%%%%%%%%%%%%%%%%%%%%%%%%%%%%%%%%%%%%%%%%%%%%%%%%%%%%%
\section*{References}
%%%%%%%%%%%%%%%%%%%%%%%%%%%%%%%%%%%%%%%%%%%%%%%%%%%%%%%%%%%%

%%%%% references %%%%%%%%%%%%%%%%%%%%%%%%%%%%%%%%%%%%%%%%%%%
\end{document}